\input psfig
\documentstyle[11pt]{article}
\renewcommand{\baselinestretch}{1.5}       
\begin{document}        
\title{Fully Relativistic Quark Models for Heavy-Light Systems}       
\author{S. A. Pernice\\         
Department of Physics and Astronomy \\         
University of Rochester, Rochetster, NY, 14627-0171}
\maketitle

e-mail: sergio@charm.pas.rochester.edu

\begin{abstract}
A fully relativistic quark model is constructed and applied to the study of
wave-functions as well  as the spectrum of  heavy-light mesons.  The free
parameters of the model are a constituent quark mass and (on the lattice)
an adjustable r-parameter in the fermionic kinetic energy, while the
confinement is introduced via potentials measured by MonteCarlo. The results
are compared to Monte Carlo energies and  Coulomb-gauge wave functions. 
They are in very
good agreement with the data. A comparison with previous models 
suggests that  we are seeing in the Monte Carlo data the
quantum-relativistic delocalization of the quark due to Zitterbewegung.

\end{abstract}
\newpage
 
\section{Introduction}

Recent studies in quenched lattice QCD \cite{tony1} have led to a considerable
advance in our understanding of meson wave functions - in particular,
of the relation between the Bethe-Salpeter wavefunction of a 
heavy-light meson in Coulomb gauge QCD and the wavefunctions obtained
from a spinless relativistic quark model (SRQM) defined by a
Hamiltonian of the form \cite{tony2,erhard}
\begin{equation}
\label{eq:H1}
H_1 = \sqrt{p^2 + m^2} \; + V(r)
\end{equation}
where $m$ is a constituent quark mass, and $V(r)$ the confining
potential (determined by Monte Carlo measurements of Wilson line 
correlations of static color sources).

Wave functions obtained from
(\ref{eq:H1}) have proved to be enormously useful in constructing
appropriately smeared lattice operators for heavy-light mesons [1], 
leading to accurate lattice calculations of B-meson properties.
They have also been recently applied to the extraction of the 
Isgur-Wise function \cite{Isgur}. Relativistic potential models have also
been used to estimate pseudoscalar meson decay constants \cite{cea}

Despite the fact that SRQM wavefunctions give a vastly better fit
than nonrelativistic ones
to the meson wavefunctions measured in Monte Carlo calculations,
some persistent discrepancies in simultaneously describing
the asymptotic (large distance) behavior as well as the
wavefunction at the origin suggest that the model defined by
Eq(\ref {eq:H1}) is not capturing all of the essential physics, even at the
level of a valence quark description. Recall that the SRQM of (\ref {eq:H1})
has only a single free parameter, the constituent quark mass $m$, as the
potential $V(r)$ is determined by Monte Carlo measurements for each
lattice studied. These discrepancies are not very important
in constructing smeared operators for the ground state meson in each
angular momentum channel, but become very troublesome when one tries
to extract excited state properties using the multistate formalism
of Ref\cite{tony1}, where admixtures of the ground state should be kept to 
a minimum. 

Our objective in this paper is not only to construct an improved version
of the SRQM which does a better job in fitting the global behavior of
meson wavefunctions for different angular momenta and for small as well
as large distance, but also to provide a clear explanation of the 
approximations being done and the relation of the resulting model with a
hypothetical full QCD solution of the problem.

 The resulting model extracts, we believe, the full content of the physical
picture provided by the valence quark description and consistent with
QCD. The accurate predictions for the wave functions as compared to
Monte Carlo simulations (see Section 3.1) indicates that Heavy-light
mesons can be represented reasonably well in terms of this picture.

The two main effects which emerge from the more complete treatment
given in Sections 2 and 3 below of the lattice QCD Coulomb gauge
Hamiltonian, and which are found to improve considerably the
agreement of the model with the measured Monte Carlo wavefunctions are
\begin{enumerate}
  \item 
A renormalization of the Wilson r-parameter away from
the bare value (r=1) used in the Monte Carlo simulations.
The sign of this lattice effect can be understood already from the one-loop
seagull correction (see Section 2.1), although the magnitude (as
in the case of the quark mass correction renormalizing $K_c$) seems
to involve a large nonperturbative piece. This is reasonable,
since a renormalization of r is an effect  involving {\it all} momenta,
in particular low momenta where we know perturbation theory fails.
Also , one must keep in mind that a one-loop calculation in the 
4-dimensional Euclidean theory (with $a_t \neq 0$), will not 
necessarily give the correct quantitative shift of the spatial 
r-parameter in the Hamiltonian formulation (where a continuum
limit $a_t \rightarrow 0$ has implicitly been taken).

This effect, which should became irrelevant in the continuum limit,
plays however an important quantitative role improving the agreement
between model and data for the lattice sizes tested so far (see section 2.2).

 \item
  Some of the observed discrepancies between model $H_1$ and the Monte 
Carlo simulations  persist, even after the corrections implied in point 1.
These remaining discrepancies are considerably reduced when the correct relativistic
treatment of the heavy-light system is performed. A detailed analysis of the
differences between this correct treatment  and the previous models give
rise to a beautiful explanation of this new corrections. They turn out to be
due to the delocalization of the light quark known as Zitterbewegung, that,
as is well known, arise  from the inability to localize a relativistic particle in
a local unitary theory.
To my knowledge, these effects are seen 
for the first time in Monte Carlo measured wave functions.

\end {enumerate}

In section 2,  we construct  a  model that  correctly takes into 
account the Wilson lattice fermionic kinetic energy \cite{Wilson} used in the 
Monte Carlo simulations. This model however does not represent 
an improvement over $H_1$. The reason for that is analyzed and as a
result a new  model arises, incorporating the renormalization of 
the Wilson $r$-parameter,  that does represent an improvement over
(\ref{eq:H1}). In section 2.2 we compare this new model and $H_1$
with the Monte Carlo data. In section 3 we carry out a fully 
relativistic treatment of the problem. In section 3.1 this model is
compared with the Monte Carlo data.
In section 4 we compare the physical content of the three models
and interpret the differences. In section 5 we present the conclusions
and discuss upcoming studies.

\section{Improved Treatment of Kinetic Terms}

As  was shown in Refs. \cite{tony1,tony2},the Hamiltonian given by equation
 (\ref{eq:H1})   describes
very well the results of   Monte Carlo calculations of the Coulomb gauge
wave functions of a heavy-light meson in quenched approximation.
In addition to practical implications for lattice studies, this model
provides a surprisingly simple physical picture for the heavy-light mesons,
namely, the heavy quark acting as a source of the confining Coulomb
potential and the light quark moving relativistically in this confining
field (the relativistic nature of the kinetic energy was essential  \cite{tony2} in reproducing
the large distance behavior of the wave function). The real gluons are completely
decoupled from the quarks except for their role renormalizing the mass.

In this paper,  we will carry the physical picture implied by a
valence quark model to its limits. The resulting model highly improves the 
one given by Eq.(\ref{eq:H1}) both conceptually and in its predictive power while
keeping the underlying simplicity.

\subsection{Using the Wilson Action}

A first, perhaps obvious modification to $H_1$  amounts to replacing
the kinetic energy by the lattice Wilson dispersion relation \cite{Wilson} taking correctly
into account the specific lattice formulation employed in the simulations. 
 It is important in assessing
the quantitative validity of the relativistic quark model that systematic effects
due to lattice discretization be dealt with consistently both in the model and
in the Monte Carlo simulations so that deviations between the two may be 
properly attributed to important physical effects rather than lattice artifacts
which will eventually disappear in the continuum limit. The Monte Carlo
calculations \cite{tony1} that constitute the `experimental' data were done with a
Wilson $r$ parameter equal to one. So our new Hamiltonian becomes:
\begin{equation}
\label{eq:Hr1}
H' = \sqrt{ M^2({\bf q}) + \sum_{i =1}^{3}{{\bf Q}_i^2}} \;+\; V(r)
\end{equation}
where 
\begin{eqnarray}
\label{eq:mHr1}
M({\bf q}) & \equiv & m + \sum_{k=1}^{3} (1 - \cos q_k )  \\
\label{eq:qHr1}
{\bf Q}_{\bf k}({\bf q}) & \equiv & \sin q_k  
\end{eqnarray}

Although this model is closer to  lattice QCD since it contains the correct 
dispersion relation, the corresponding wave functions {\em do not} represent
an improvement with respect to model (\ref{eq:H1}). Actually, they  magnify
 the  discrepancies between model $H_1$ and Monte Carlo data.
This is at first sight very surprising because, as already said,  Eq(\ref{eq:Hr1}) is
closer to lattice QCD in its treatment of the fermionic kinematics  than $H_1$. 

The solution to this puzzle comes from a detailed analysis
of the renormalization of the parameters of the theory {\it  on the lattice}. 

More specifically, consider  the one loop contribution to the quark self energy. 
On the lattice we have {\em two} graphs rather than one (as a consequence of
the compact representation of the gauge field): 

\begin{figure}[htp]
\hbox to \hsize{\hss\psfig{figure=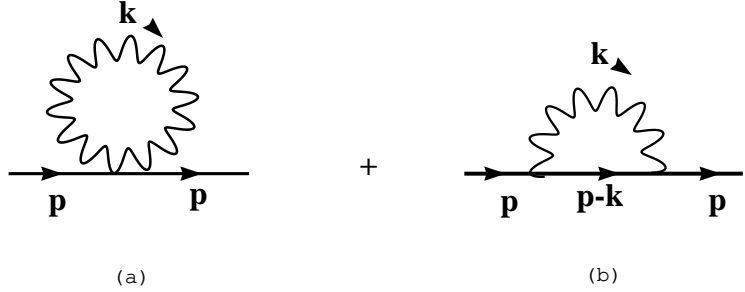,width=0.8\hsize}\hss}
\caption{One loop graphs contributing to the quark self-energy}
\label{fig:feyn}
\end{figure}

Corresponding to:
\begin{equation}
\label{eq:gamma}
\tilde{\Gamma}_p \equiv \tilde{\Delta}_p^{-1} -  \Sigma_p  = m + {r\over2} \hat{p}^2 
+ i  \gamma\cdot\bar{p}  + \Sigma_p^{(a)}  + \Sigma_p^{(b)}
\end{equation}
where $ \hat{p} = 2 \sin{p_\mu\over2}$ and $\bar{p}_\mu = \sin{p_\mu}$.\footnote{We are
using here the notation of Ref\cite{lattbook}}

Graph (b)  also appears in the continuum while graph (a) is present only on the lattice
in a compact formulation of the gauge theory.
It is precisely graph (a)  that  will provide in the cleanest way the solution
to our puzzle, as its contribution to the self energy  in Coulomb gauge is:
\begin{eqnarray}
\label{coulomb}
\Sigma_p^{(a)}  &=&   g^2 {(N^2 - 1)\over 4N}  \left[ {1\over\Omega} \sum_k  {1\over
 \hat{k}_i^2 } \right] 
(r \cos{p_0} - i \gamma_0 \bar{p}_0)  + \\
\label{gluons}
 & & g^2 {(N^2 - 1)\over 4N} \sum_{i = 1}^3 \left[ {1\over\Omega}  \sum_k  {1\over
 \hat{k}_{\mu}^2 }  ( 1 - {\hat{k}_i^2 \over | \hat{k}_j^2| } ) \right]
(r \cos{p_i} - i \gamma_i \bar{p}_i) 
\end{eqnarray}
where $\Omega = L^4$, greek indices run from 1 to 4 and roman indices from 1 to 3
(this convention applies in all equations in this paper)
. Eq(\ref{coulomb})contains the contribution from the Coulombic instantaneous interaction
while Eq(\ref{gluons}) contains the contributions from the real gluons.
 
 Writing $\hat{p}^2$ as $ \sum_{\mu = 1}^{4}2 (1 - \cos{p_\mu} )$, the inverse
free propagator becomes
\begin{equation}
\label{invprop}
 \tilde{\Delta}_p^{-1} = m + 4 r  - r \sum_{\mu = 1}^{4}\cos{p_\mu} 
+ i  \gamma\cdot\bar{p} 
\end{equation}
and we immediately realize that the part  of  $\Sigma_p^{(a)}$ proportional to the identity
matrix (in the Dirac indices)  explicitly renormalizes the Wilson $r$ parameter.
Specifically:

\begin{eqnarray}\label{rtime}
r_{{\rm time}} & \rightarrow  &r \left\{ 1 -  g^2 {(N^2 - 1)\over 4N}  \left[ {1\over\Omega} \sum_k  {1\over
 \hat{k}_i^2 } \right]  \right\}  \\ \label{rspace}
r_{{\rm space}} &\rightarrow  &r  \left\{ 1 -  g^2 {(N^2 - 1)\over 4N}  \left[ {1\over\Omega}  \sum_k  {1\over
 \hat{k}_{\mu}^2 }  ( 1 - {\hat{k}_i^2 \over | \hat{k}_j^2| } )  \right]  \right\}
\end{eqnarray}
For our lattice size, the perturbative $r$ renormalization 
due to graph (a) are, in Coulomb gauge:
\begin{eqnarray}\label{deltartime}
\delta r_{{\rm time}}^{V = 12^3}  &=&   -  g^2 {(N^2 - 1)\over 4N}  \; 0.234 = -  0.452\\
\label{deltarspace}
\delta r_{{\rm space}}^{V = 12^3} &=& -  g^2 {(N^2 - 1)\over 4N}  \;  0.102 = -  0.197
\end{eqnarray}

We shall be comparing RQM models with MonteCarlo data generated on a 12$^3$x24
lattice at $\beta=$5.7, corresponding to a naive bare lattice coupling $g_0^2\sim$ 1.05.
The hopping parameter was $\kappa =$ 0.168.
Nonperturbative effects may partially be included by using instead the tadpole-improved \cite{LepMack} definition of the coupling, which gives for the $\beta$ value considered
a value closer to 2.9 for $g^2$ \cite{tony2}.  This is the value used in 
Eqs(\ref{deltartime},\ref{deltarspace}).

In our Hamiltonian models we consider of course only  $r_{{\rm space}}$. 
This value, as we will see in the next subsection correctly predict the sign of the 
change in $r$ although the  magnitude seems to  have big nonperturbative contributions.
Graph (b) also contributes {\em effectively}  to the  $r$ renormalization, but
not in an explicit way as in the case of the first one. However, in this case the numerical
contribution is much smaller (as in the case of the mass shift).

Of course  the mass is also  renormalized as  is well known, and also by an amount which is
quite a bit larger than the perturbative one-loop shift (even with tadpole improved
couplings).

The important point of this calculation is to realize that  {\em not only the mass
but also the Wilson $r$ parameter  should be considered as free parameters, since both
of them are dynamically modified, in a nonperturbative way}.

Including this effect,  the model acquires the same form as in Eq(\ref{eq:Hr1})
\begin{equation}
\label{eq:H2}
H_2 = \sqrt{ M^2({\bf q}) + \sum_{i =1}^{3}{{\bf Q}_i^2}} \;+\; V(r)
\end{equation}
but with 
\begin{equation}
\label{eq:H2mass}
M({\bf q})  \equiv  m + r\sum_{k=1}^{3} (1 - \cos q_k )
\end{equation}
We have now therefore  {\it two} adjustable parameters, $m$ and $r$. This new model,
with correctly chosen values for the parameters, represents a  substantial quantitative
improvement over model (\ref{eq:H1}) as will be shown in the next section. We
also understand now why Eq.(\ref{eq:Hr1})  actually works worse than Eq(\ref{eq:H1}),
 as $H_1$ is {\it effectively} close (in the sense that the fermionic kinetic dispersion 
relation is close to the bosonic one over most lattice momenta)
to one particular case of the model $H_2$. In fact, it corresponds, for fixed
$m$, to $r\approx 0.85$ as can be seen simply by plotting the corresponding
dispersion relations. This value, although not  optimal, is closer
to the optimal choice for model (\ref{eq:H2})  (see Section 2.2) than  the naive unrenormalized choice
$r=1$. 

The  improvement  obtained with Eq (\ref{eq:H2}),  although 
very significant from a quantitative point of view
for the lattice sizes tested so far ,  should nevertheless become irrelevant
in the continuum limit, although it is certainly relevant in providing accurately
smeared meson operators for multistate MonteCarlo studies \cite{tony1}. 

 In any case, we have now not only a better model
but  one that has a closer connection to QCD since it  contains the
actual  {\it dynamical}  QCD fermionic kinetic energy .  We shall see in the
next  section that the modification in the dispersion
formula greatly improves the fit to the measured wavefunctions at shorter
distance (and in particular at the origin) once the $m$ and $r$
parameters are chosen to optimize the fit at medium and large distances.

   A fuller description,
starting with the Bethe-Salpeter equation (which for a light quark propagating
in  the color field of a static source reduces to a Dirac equation) will lead  in 
Section (3) to  a model giving 
similar wavefunctions, agreeing even more closely with the measured ones.
Such a model represents  a valence quark description of
the heavy-light meson that is as close to QCD as possible without
leaving the physical picture outlined in the introduction.

\subsection{Quantitative Consequences of the Improved Potential Model}

In order to actually solve for the wave functions of the model, we used the same method
as in Refs \cite{tony1,tony2}.  We briefly explain it   here for completeness.

  The procedure used in a multistate smearing calculation of heavy-light meson
properties \cite{tony2} for generating lattice smearing functions from the
RQM is as follows. One obtains orthonormal lattice wavefunctions,
which are eigenstates of a lattice RQM Hamiltonian defined on a
$L^{3}$ lattice (with $\vec{r},\vec{r}^{\prime}$ lattice sites):

\begin{eqnarray}
\label{Hnum}
  H_{\vec{r}\vec{r}^{\prime}}&\equiv& K_{\vec{r}\vec{r}^{\prime}}
  +V(\vec{r})\delta_{\vec{r}\vec{r}^{\prime}}   
\end{eqnarray}

The eigenstates in a channel of given orbital quantum numbers (S,P,D etc)
are obtained by applying the resolvent operator $\frac{1}{E-H}$ to a source
wavefunctions of the same orbital symmetry. The model at this stage is
spinless (the measured wavefunctions represent spin-averages of the top
two Dirac components of the light quark field) so issues of spin-orbit coupling
do not yet arise (they will be dealt with properly in the full Dirac formalism
of Section 3).

  In the resolvent approach, S-states are generated by applying the resolvent kernel
 to a monopole localized at the origin, P-states with a source dipole, and so on.
 At each trial value of the energy $E$, the norm of the resulting state $\frac{1}{E-H}\Psi^{(0)}$
is evaluated. Obviously
\begin{equation}
 R\equiv \|\sum_{r^{\prime}}(\frac{1}{E-H})_{\vec{r}\vec{r}^{\prime}}
 \Psi^{(0)}(\vec{r}^{\prime})\|\rightarrow\infty
 \end{equation}
when $E\rightarrow {\rm eigenvalue \;of}\, H$.
Typically, wavefunctions accurate to 4-5 significant figures are
obtained by stopping once this norm exceeds 3000. At this point
a smearing eigenstate $\Psi^{(a)}_{\rm smear}(\vec{r})$ is extracted
by renormalizing the vector $\frac{1}{E-H}\Psi^{(0)}$ to unit norm. The
inversion of $E-H$ is performed by the conjugate gradient algorithm,
with the multiplication of the kinetic term done in momentum space
using a fast Fourier transform.

In the following figures we present  the results of our new model as
compared with the old one. As was already mentioned in the 
previous section, the MonteCarlo data presented in the foolowing
figures was generated on a 12$^3$x24 lattice at $\beta=$5.7, and
the hopping parameter was $\kappa =$ 0.168.

\begin{figure}[htp]
\renewcommand{\baselinestretch}{1.0}
$$\psfig{figure=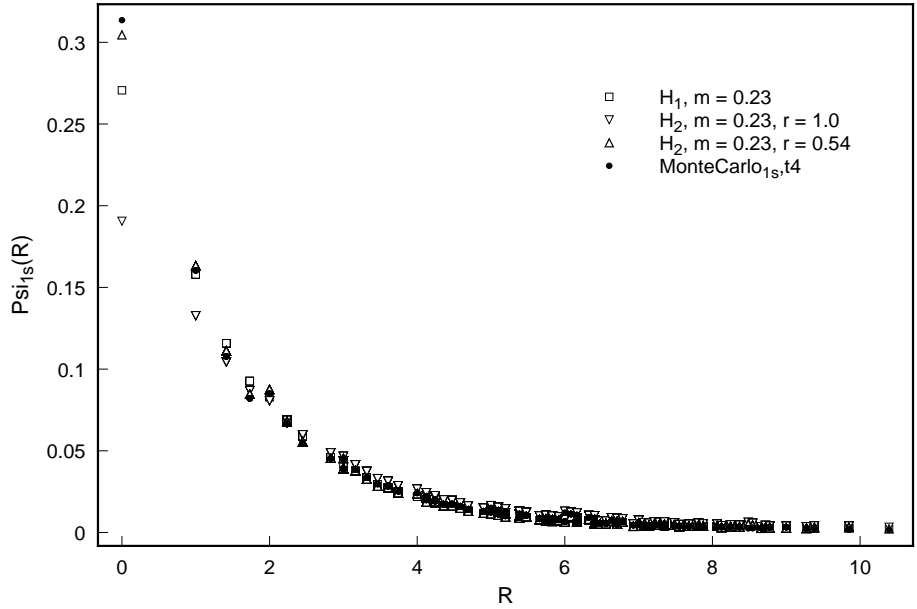,width=1.0\hsize}$$
\caption{ 1S state, $L = 12$. We see at the origin the improvement of  $H_2$ over $H_1$
when $r$ is renormalized. With $r = 1$ however, model $H_2$ does a poor job
showing the necessity of  $r$ renormalization. The Monte Carlo wave functions were extracted
at different time slices. Although all time slices gave very similar results, the wave function
extracted at the fourth one, that we present here, was the one with the best
signal to noise ratio. That is  the meaning of  the t4 in the Monte Carlo data point label.}
\label{fig:s1L12}
\end{figure}

In Fig[\ref{fig:s1L12}] we compare the
ground state wave functions for the $ V = 12^3 $ case. The values
chosen for the constituent mass and $r$ parameters are  chosen to
maximize the agreement (in a mean square sense) between data and
the respective models {\em in the ground state}. As it turned out, the optimum
constituent masses are very similar to one another and the wave
functions very insensitive to small changes around the optimum 
value. We present here the results for  the same values of the constituent mass.
This choice, while essentially identical to the optimum cases, helps to appreciate
the effect of the r renormalization. The case of $H_2$
with $ r = 1 $ is also included to emphasize the effect of  the $r$
renormalization.  As we can see, the agreement  with the Monte Carlo
data  was  already very good for $H_1$ and is further improved,
specially at the origin by $H_2$.

\renewcommand{\baselinestretch}{1.5}

\begin{figure}[htp]
$$\psfig{figure=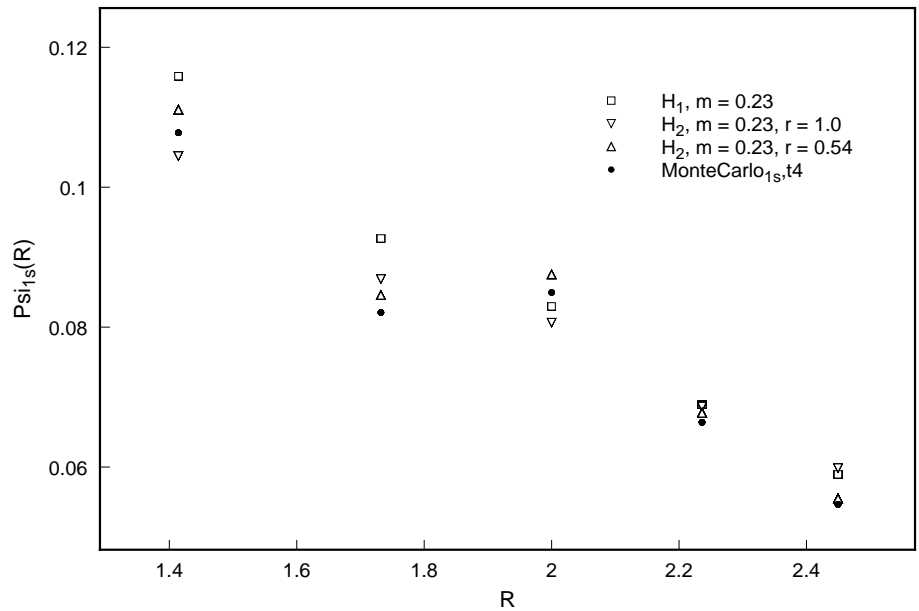,width=1.0\hsize}$$
\caption{  Detail of figure 2. We see here the `discontinuity' between 
points at  distances $R_1 = 1.73 $ and $R_2 = 2$. While model $H_1$ completely
ignores it, and model $H_2$ with unrenormalized $r$ can do just slightly better,
model $H_2$ with the renormalized $r$ almost perfectly follows the discontinuity. }
\label{fig:s1L12detail}
\end{figure}

But the most important  reason for which model $H_2$ was introduced, was
to capture the lattice artifacts unavoidably present  in the Monte Carlo
data. Only after these artifacts are well under  our control  can we hope to find
some physics in the data beyond the one provided by $H_1$. In this sense
the improvement at the origin is due to the $r$ renormalization as can be seen
by comparing with the unrenormalized case denoted $H_2$, $m = 0.23$, $r = 1.0$
Also we present in Fig[\ref{fig:s1L12detail}] a detail of  Fig[\ref{fig:s1L12}]
corresponding to the region of distances between  $R = 1.4$ and $R = 2.4$. 
Specifically, as can be seen in Fig[\ref{fig:s1L12}], at points corresponding to distances
 $ R_1 = \sqrt{ 3} $ ( this corresponds to the lattice points $ \vec{x}_1 = \pm  1 \hat{i}
 \pm 1 \hat{j} \pm 1 \hat{k} $ ) and  $R_2 = 2$ (corresponding to the point
$ \vec{x}_2 = \pm 2 \hat{i} \pm 0 \hat{j}  \pm 0 \hat{k} $ and the points generated
by cyclic permutation of the  coordinates), there is a pronounced `discontinuity'
 in the Monte Carlo data  that  should of course disappear  in the continuum limit.
 On a finite lattice and still not very close to the continuum 
this discontinuity is easy to understand qualitatively: it is due to the fact that  under these 
conditions the system responds more naturally in terms of  a metric notion of distance between 
two points on a lattice given  by some function of the number of  links between  
these points (notice that  $ \vec{x}_1$ is at 3 links away from the origin while 
$ \vec{x}_2$ is only at 2, in contradistinction with their euclidean distance).
In figure \ref{fig:s1L12detail} we see how model $H_1$ completely ignores this lattice
artifact, $H_2$ with $r = 1$ is slightly closer, while $H_2$ with $r = 0.54$ follows
almost perfectly  the discontinuity.

\renewcommand{\baselinestretch}{1.5}

\begin{figure}[htp]
\renewcommand{\baselinestretch}{1.0}
$$\psfig{figure=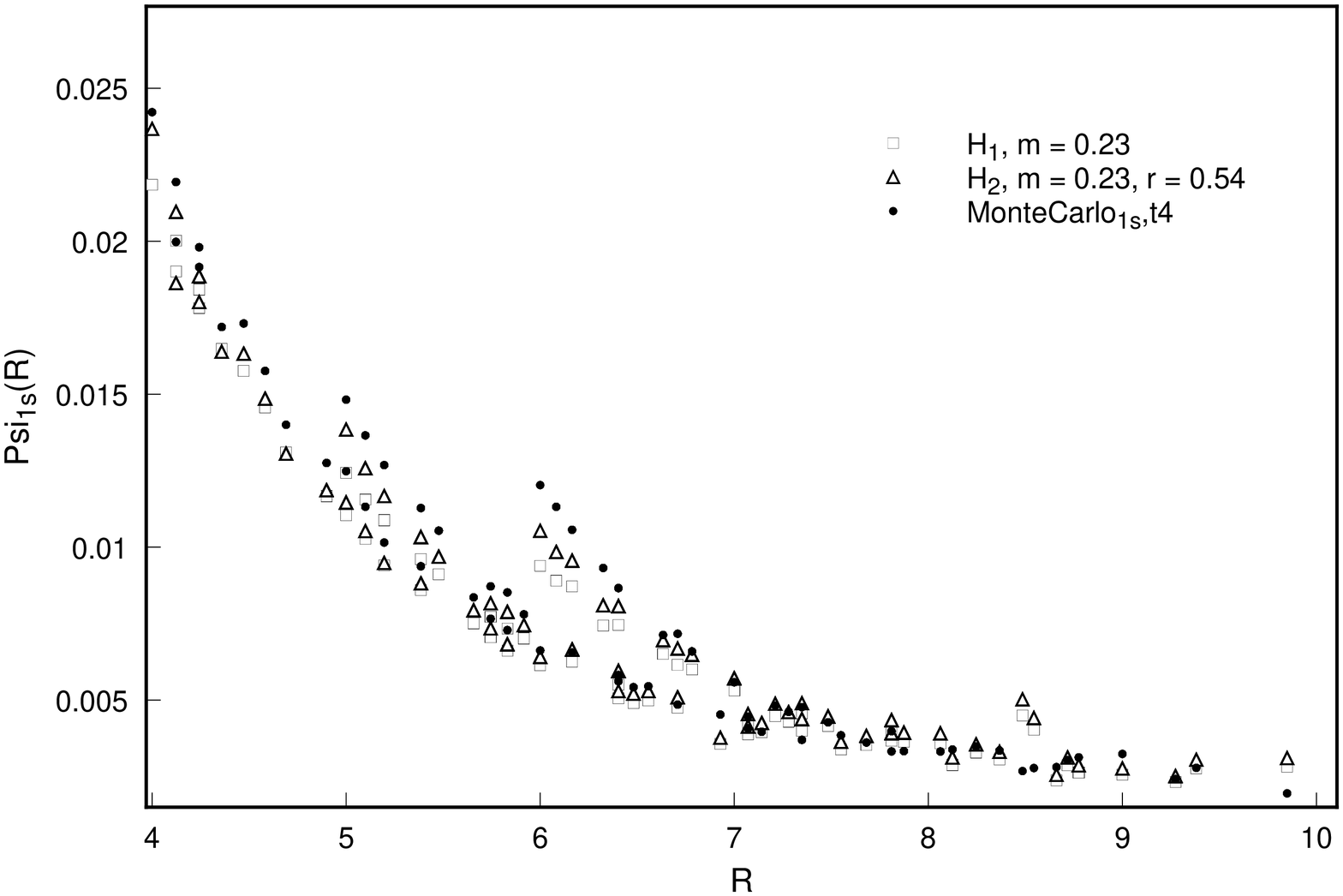,width=1.0\hsize}$$
\caption{ Large R region of figure 2. The case $H_2$ with $r = 1.0$ is not
displayed to clarify the relevant information. We see that both $H_1$ and $H_2$ with
renormalized $r$ fall very close to the data in this region (notice the scale). }
\label{fig:s1L12largeR}
\end{figure}

In Fig[\ref{fig:s1L12largeR}] we can better appreciate the large distance region.

\renewcommand{\baselinestretch}{1.5}

\begin{figure}[htp]
\renewcommand{\baselinestretch}{1.0}
$$\psfig{figure=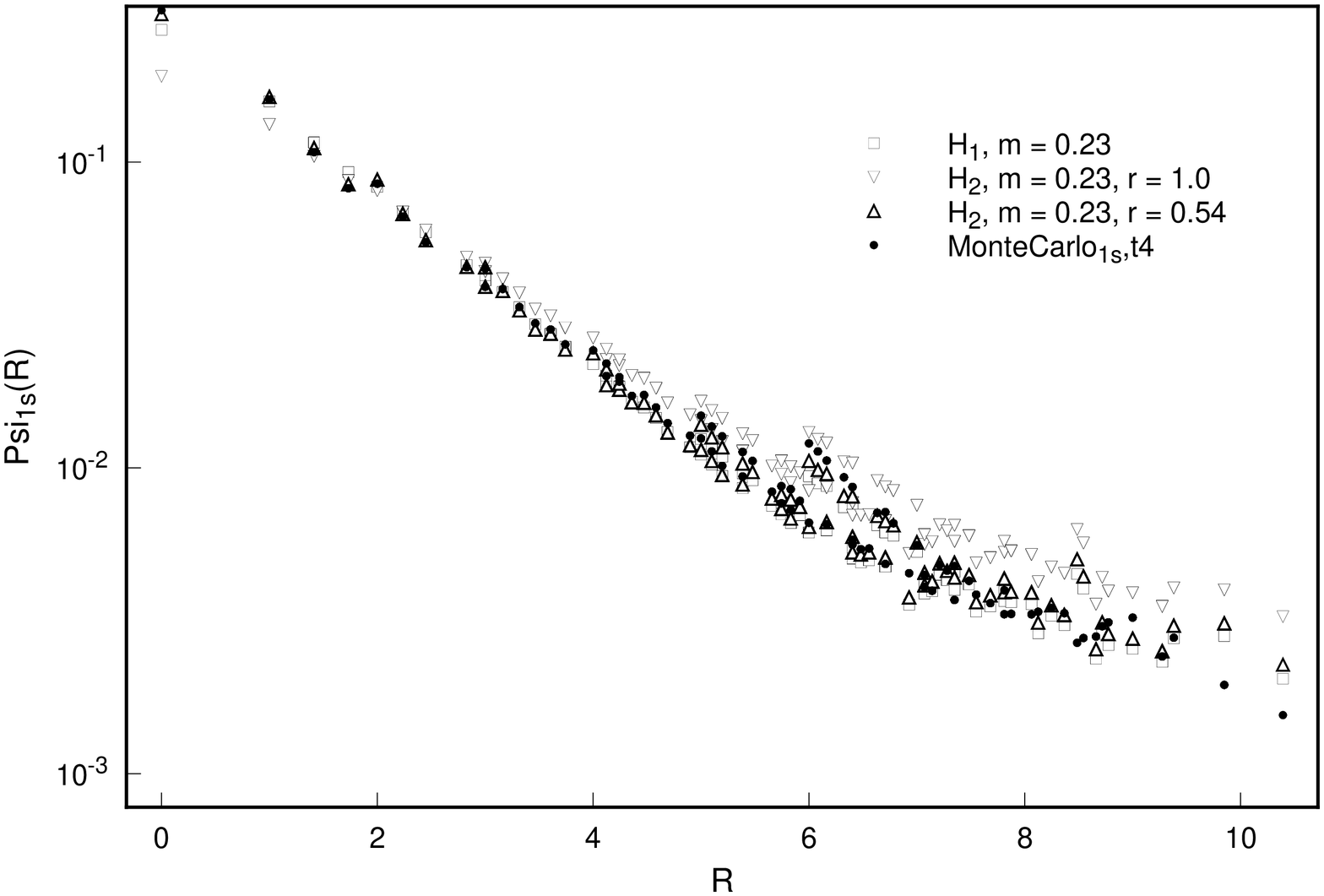,width=1.0\hsize}$$
\caption{1S state $L = 12$ , logarithmic scale. Both, $H_1$ and $H_2$ with
renormalized $r$ do a very good job at large distance.}
\label{fig:logs1L12}
\end{figure}

In Fig[\ref{fig:logs1L12}], we show the same information as in Fig[\ref{fig:s1L12}]
 but in logarithmic scale
to appreciate the asymptotic region. As we see, both, $H_1$ and $H_2$ with
renormalized $r$ do a very good job in this region.

\renewcommand{\baselinestretch}{1.5}

So, as we have seen, as far as the ground state is concerned,  $H_2$ not only shows
an improvement  over $H_2$ specially visible at the origin, but it  also proved
capable of capturing very pronounced lattice artifacts. Both effects clearly 
show the relevance of taking into account the $r$ renormalization.

\begin{figure}[htp]
\renewcommand{\baselinestretch}{1.0}
$$\psfig{figure=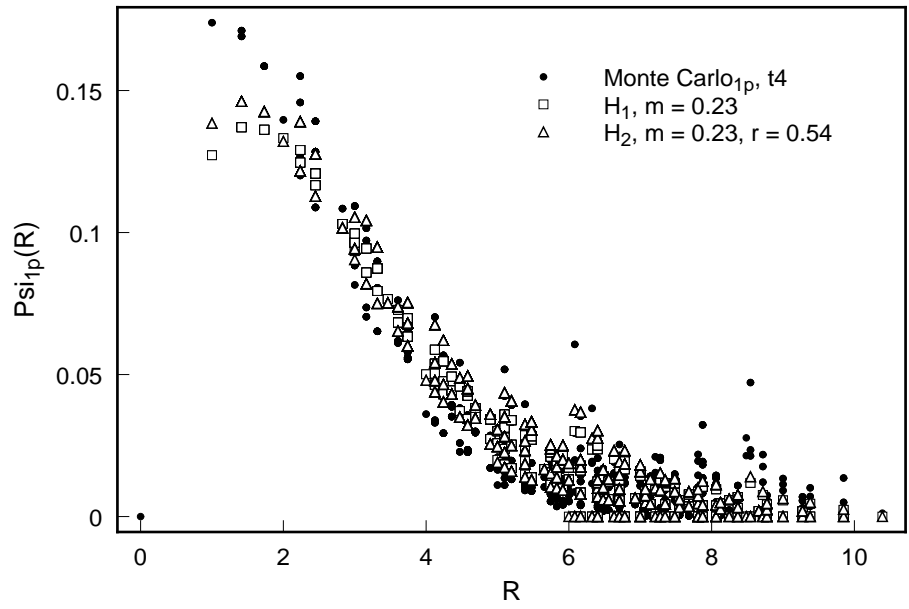,width=1.0\hsize}$$
\caption{1P state , $L = 12$. The values of the parameters were fixed to reproduce
as accurately as possible the ground state. We can see the improvement of 
$H_2$ over $H_1$, but still we have plenty of room to improve.}
\label{fig:p1L12}
\end{figure}

Once the values of  $m$ and $r$ are specified to reproduce as accurately as possible 
the ground state, we compare now the results for the 1P state. In this case,
we divide the  respective wave functions by $\cos{\theta}$ to show only
the  radial dependence. As we can see in 
Fig[\ref{fig:p1L12}], the model $H_2$ does again a better job than $H_1$, although
there is still  room to improve. The case of $H_2$ with $r = 1$  is not shown
since it  was in the previous figures only to see the effect of renormalizing $r$. In any
case, it again performs worse than $H_1$. 

\renewcommand{\baselinestretch}{1.5}

So we conclude that, although the modifications leading to $H_2$ are only 
due to lattice artifacts,
the quantitative improvement is significant, so the value of $H_2$ resides in the fact
that it captures a very important lattice discretization effect.  Nevertheless the 
improved model is still conceptually and quantitatively inadequate. The conceptual
inadequacy stems from the fact that the relation between the eigenstates of $H_2$
and the spin-averaged Bethe-Salpeter wavefunctions in Coulomb gauge is unclear
(e.g. the potential model ignores antiquarks whereas there are coupled 
upper and lower components in a Dirac formalism).
Quantitatively, we shall see that use of a full Dirac formalism which is closely 
related to the Bethe-Salpeter wavefunction also further improves the agreement with the
Monte Carlo results. In this full formalism, it will still be important however to include
the $r$-renormalization discussed above.

\section{Full Bethe-Salpeter treatment of  Heavy-Light Wavefunctions}

As we have seen, the agreement between the wavefunctions derived from
the Hamiltonian $H_2$  and the
Monte Carlo data  is quite remarkable; however, not only is there still room for 
further quantitative improvement but  from a conceptual point
of view  the connection between these simple models and a full hypothetical QCD 
solution of the meson Coulomb gauge wave functions is not completely 
clear. In another words, it would be nice to have a model that works as
well as the previous one and in which the nature of the approximations being
done is completely transparent. In this subsection we will construct this model and as a 
bonus the resulting one will show an additional quantitative improvement over $H_2$ with 
a very nice physical interpretation.

 We shall assume that: \\
(a) Transverse gluon interactions with the quarks act primarily to renormalize
the mass and r parameters in the quark kinetic term. Fock states involving real gluons
in addition to the valence quarks are neglected.\\
(b) The net effect of Coulomb gluon exchange between the light and static 
quarks can be expressed by the  potential acting between two
infinitely heavy color sources.

More qualitatively, the picture in the back of our mind, supported by the comparison
with data as will be seen in Section (3.1), consists of the light quark moving fast enough
for  relativistic effects to be important, but on the other hand not so fast that the 
interaction with the static quark cannot be accurately described by the energy
which would obtain if the light quark were held fixed. Alternatively, one might assume that
the time scales over which the string connecting the quark to the static source
responds to changes in the light quark position are small compared with the time
scales relevant for the light quark motion.

Before we proceed with the derivation of our new model, it will be useful to 
present a brief  description of what  was  actually  measured in the 
Monte Carlo simulations of Ref\cite{tony1} that constitutes our data. Even though 
this work used a sophisticated multistate smearing method,  for our purposes it
suffices to know that the basic information was extracted from the measurement,
in quenched lattice QCD,  of  the Green function: 
\begin{equation}\label{MCgreenfct}
F(\vec{x}' , \vec{x}, t ) =  \langle 0| \bar{Q}_H (\vec{0},t) \gamma_5 q_H(\vec{x}',t) 
\bar{q}_H(\vec{x},0) \gamma_5 Q_H (\vec{0},0)|0 \rangle
\end{equation}
in the limit were the b-quark is taken to be infinitely massive. In this limit, the heavy
quark propagator  is simply proportional to $\frac{1 - \gamma^0}{2}$, therefore $F$
becomes proportional to the average of the upper two components of the light quark propagator
in the presence of  a  color source.
 From the calculation of this object, using the above
mentioned multistate smearing method  (i.e. smearing the source point
$\vec{x}$ of the light  quark with  Ansatz meson wavefunctions derived from
$H_1$)   the upper two components of the meson wave function
were extracted and spin averaged.
 The result of this operation constitutes the data against which  we compare our models.

Taking this into account  we will now construct  a model that  represents as 
closely as possible the quantities  measured   in the Monte Carlo simulations
realizing at the same time the physical ideas presented above.  

In a full QCD treatement  of the problem at hand, the relevant Bethe-Salpeter
wavefunction would be 
\begin{equation}
\label{eq:fullBS}
\chi(\vec{x},t) \equiv \langle 0|q_H(\vec{x},t)\bar{Q}_H (\vec{0},t)|P\rangle
\end{equation}
where $|0\rangle$ is the vacuum, $|P\rangle$ is the meson state (in the center of
mass frame with energy $H|P\rangle=E_{BS}|P\rangle$), 
 and $q_H ,  Q_H$ are the light and heavy Heisenberg fields.

In the infinitely massive heavy quark limit, but otherwise still in full QCD,
Eq.(\ref{eq:fullBS}) is best written as, 
\begin{equation}
\label{eq:BSHeavyM}
\chi(\vec{x},t) \equiv \langle 0|q_H(\vec{x},t)|P * \rangle
\end{equation}
where $|P * \rangle \equiv \bar{Q}_H (\vec{0},t)|P\rangle$.  This notation emphasizes the
fact that in the above limit, the heavy quark field is not dynamical.

As we see, if  we were able to calculate exactly  Eq.(\ref{eq:BSHeavyM})  in the context of 
heavy quark limit quenched lattice QCD, we would be reproducing every detail of  the 
results of the Monte Carlo simulations, since that is precisely the quantity being measured.

In our physical picture however, as stated above the 
transverse gluon interactions with the quarks act primarily to renormalize
the mass (and in the lattice also the Wilson $r$ parameter) in the quark kinetic term
and  the net effect of Coulomb gluon exchange between the light and static 
quarks can be expressed by the  potential acting between two
infinitely heavy color sources. Under these conditions the equation satisfied by
$q_H$ reduces to:
\begin{equation}\label{modeleq}
\langle 0 |(  \gamma_E^0 {\partial \over \partial t} -i \vec{\gamma}\cdot\vec{\nabla}
+ i \gamma_E^0 A_0  + m ) q_H |P * \rangle = 0
\end{equation}
that  together with the Heisenberg equation ${\partial \over \partial t}{q_H = [H,q_H]}$
(in Euclidean space) 
and the relation $\langle0|[H,q_H]|P * \rangle = -E_{BS}\langle0|q_H|P * \rangle$, give
rise to the eigenvalue equation
\begin{equation}\label{dirac}
( -i\vec{\alpha}\cdot\vec{\nabla} + m \beta + V(\vec r)) \chi(\vec r)  =  E_{BS} 
\chi(\vec r)
\end{equation}
which is nothing but the Dirac  equation  for the light  quark in the 
presence of the confining external field . This equation corresponds, on the lattice,
(with the renormalization of the Wilson $r$ parameter  also  taken into account)
to an effective lattice Hamiltonian given by the usual Wilson fermion action:
\begin{eqnarray}
H_3 & = & \sum_{\bf x} \{ \chi^{+} ({\bf x}) ( m + 3r ) \; \beta \; \chi ({\bf x})  
\nonumber  \\
 &   &  + \frac{i}{2} \sum_{k=1}^{3} [ \chi^{+} ({\bf x} + \hat{\bf k})  \;
\alpha_{k}  \;  \chi ({\bf x}) - \chi^{+} ({\bf x}) \; \alpha_{k} \; \chi  ({\bf x} + \hat{\bf k}) ]   \nonumber  \\
 &   &  - \frac{r}{2}    \sum_{k=1}^{3} [ \chi^{+} ({\bf x} + \hat{\bf k}) 
 \; \beta \; \chi ({\bf x}) + \chi^{+}({\bf x}) \; \beta \; \chi ({\bf x} + \hat{\bf k}) ] 
\nonumber   \\ 
 &   &  +  \chi^{+} ({\bf x})\; V({\bf x}) \; \chi ({\bf x}) \}
\label{eq:H3}
\end{eqnarray}
where ${\bf x}$ represents a point in the three dimensional lattice of size
$L$ , $\beta$ and $\alpha_{k}$ are just the Dirac matrices,  $\chi ({\bf x})$
is the 4-component  wave function, and $V({\bf x})$ is the confining potential
 determined by Monte Carlo measurements of Wilson line correlations of
static color sources \cite{tony1}. The constituent mass $m$ and the Wilson $r$ parameters
are free parameters.

The Hamiltonian $H_3$ defines our  new model. From the above discussion
 we realize it  represents the closest possible model to QCD consistent with the valence
quark picture whose validity in the heavy-light meson system we want to check.

As will be shown in Section (3.1) this new model  represents a further improvement 
in the prediction of the correct wave functions, that by now are, within the errors
of the Monte Carlo calculations, essentially fully reproduced, indicating  the 
validity of the valence quark model to describe heavy-light  mesons. Given
the necessary assumptions to generate this picture from QCD (stated above), 
the  strong coupling nature of  the confining mechanisms, and the lightness
of one of the quarks clearly reflected in the necessity of a fully relativistic
kinetic energy, the  success of the model can hardly  be
expected a priori, and constitutes a strong statement about QCD dynamics.

\subsection{Comparison with data}

To  find  numerically the eigenvectors and eigenvalues of  $H_3$, although we followed 
in general the same procedure outlined in Section(2.2),  some features of $H_3$
had to be taken into account. For example, due to the non-positivity of  the spectrum,
the inversion of $E - H_3$ was performed with a generalization of the conjugate
gradient algorithm, the so called {\em minimum residual algorithm} \cite{numrec},
 that takes care of  symmetric but non-positive definite matrices (one may replace the
$N \times N$ complex hermitian $H_3$ with a real symmetric  $2N \times 2N  $ 
version, which is however non-positive-definite). To locate the correct  region of the
spectrum we started in the large mass regime where the wavefunctions are well
understood and gradually reduced the mass while tracking the resulting eigenstates.

In the following figures we present the results of our new model and compare them
with the Monte Carlo Data and the predictions of  $H_2$. The values of the parameters
are chosen again to reproduce as well as possible  the ground state of the system.
 Following as closely as possible what was done in the Monte carlo simulations,
(briefly described in section 3), the results of  $H_3$ presented in the figures, constitute
the average of the two upper components of the corresponding four-component
eigenvectors.

\begin{figure}[htp]
\renewcommand{\baselinestretch}{1.0}
$$\psfig{figure=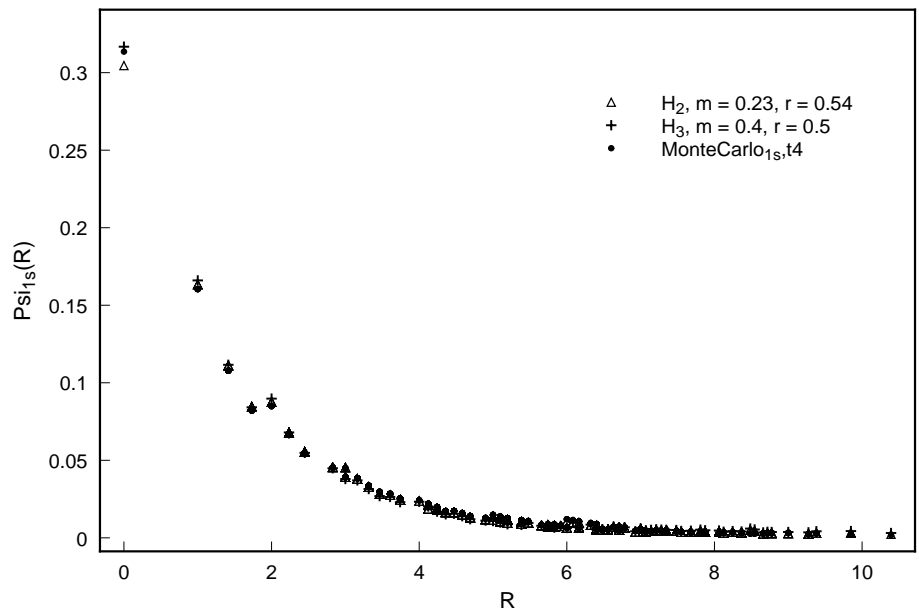,width=1.0\hsize}$$
\caption{ 1S state, $L = 12$. The Dirac model performs in this case slightly 
better than $H_2$, although  there is little room for
further improvement in this case.}
\label{fig:1SL12MCvsDvsRQM}
\end{figure}
\renewcommand{\baselinestretch}{1.5}
 
In Fig[\ref {fig:1SL12MCvsDvsRQM}] we see that our new model performs as well as 
$H_2$ for the ground state, where there was essentially no room for further improvement.
We should however note that  while the optimum value for $r$ in $H_3$ suffers only a
small change with respect to the one in $H_2$, the optimum mass becomes
considerably heavier.

Once the parameters  have been fixed to reproduce as well as possible 
the ground state of the system, we may compare the 1P state.
Again, as in the previous figures for 1P wave functions, we divide
them by $\cos{\theta}$ and present only the radial part.
In this case we clearly see the quantitative superiority of $H_3$
over the previous models. Near the origin $H_3$ falls much closer to the data
than $H_2$.

\begin{figure}[htp]
\renewcommand{\baselinestretch}{1.0}
$$\psfig{figure=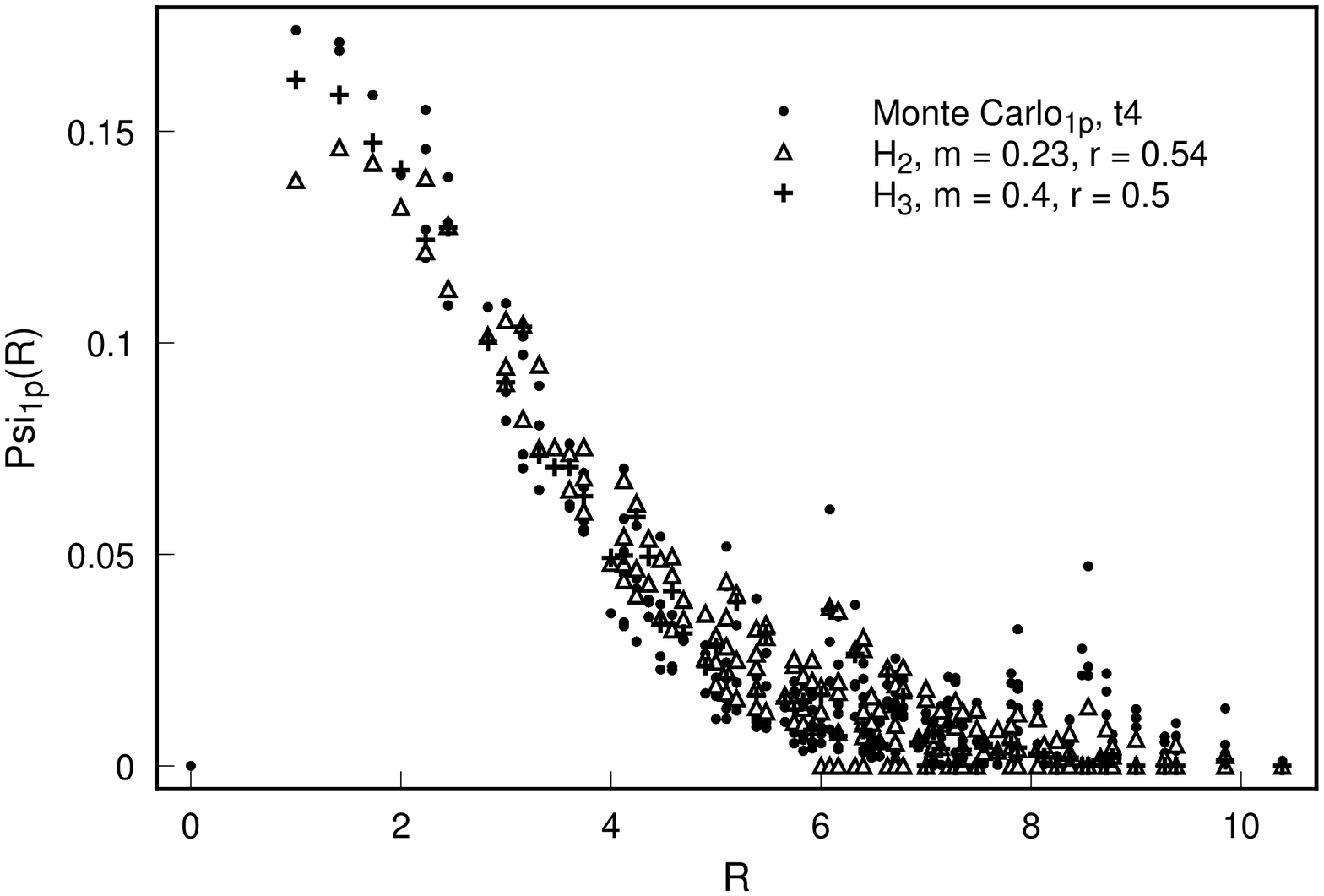,width=1.0\hsize}$$
\caption{1P state, $L = 12$. The Dirac model performs in this case much 
better than $H_2$.}
\label{fig:1PL12MCvsDvsRQM}
\end{figure}
\renewcommand{\baselinestretch}{1.5}

We see then that choosing the optimal parameters for  the respective models,
a full Dirac model based on the operator $H_3$ (that, as we have seen in section 3 is conceptually as close to lattice QCD as possible within the valence quark model),
outperforms all the other models and within Monte Carlo errors  essentially fully
reproduces the data.

We had also available the energies of the 1S and 2S states for the Monte Carlo 
data, obtained from the multistate smearing analysis of  \cite{tony1}. The only meaningful  comparison is between energy differences since
there is  an arbitrary choice in deciding the zero energy of the potential
$V(r)$. The respective energy differences between 1S and 2S states are
presented in  Table 1.

\begin{table}
\renewcommand{\baselinestretch}{1.0}
\centering
\caption{}
\vspace{.1in}
\label{tab:scales}
\begin{tabular}{|| l | l||}
\hline
  Model& $E_{2S} - E_{1S}$  \\
\hline
 Monte Carlo, $\kappa = 0.168$ & 0.31 $\pm$ 0.02  \\
\hline
 $H_1$, $m = 0.23$ & 0.381  \\
\hline
 $H_2$, $m = 0.23$, $r = 0.54$ & 0.356  \\
\hline
 $H_3$, $m = 0.4$, $r = 0.5$ & 0.324  \\
\hline
\end{tabular}
\end{table}
\vspace{0.2in}
\renewcommand{\baselinestretch}{1.5}

Again model $H_3$ is in better agreement with the Monte Carlo results 
than the others and, within the errors, reproduces the measured results.

Model $H_3$ was systematically closer to the data for other values 
of the hopping parameter $\kappa$. We present in Table 2 the energy
splitting for the Monte Carlo data corresponding to  $\kappa =  0.161$.
This value corresponds to a heavier light quark and the optimum values
of the parameters correspondingly change. They are also presented in
Table 2.  Although the Monte Carlo predictions for the various values
of the energies change with respect to the previous ones, the energy
difference essentially remains unchanged.
 This behavior is closely followed by $H_3$ that continues matching
the data. Very interestingly though, $H_2$  suffers an appreciable
modification  in the right direction, it's predictions approach the ones
of  $H_3$ for this heavier case.  The approach of models 2 and 3
for heavier quarks will be dsicussed in greater detail 
 in the following
section.

\begin{table}
\renewcommand{\baselinestretch}{1.0}
\centering
\caption{}
\vspace{.1in}
\label{tab:scales}
\begin{tabular}{|| l | l||}
\hline
  Model& $E_{2S} - E_{1S}$  \\
\hline
 Monte Carlo, $\kappa = 0.161$ & 0.32 $\pm$ 0.02  \\
\hline
 $H_1$, $m = 0.32$ & 0.385  \\
\hline
 $H_2$, $m = 0.32$, $r = 0.46$ & 0.338  \\
\hline
 $H_3$, $m = 0.5$, $r = 0.45$ & 0.325  \\
\hline
\end{tabular}
\end{table}
\vspace{0.2in}
\renewcommand{\baselinestretch}{1.5}

 In the next section we will discuss the nature of the improvement 
of  $H_3$ with respect to the previous models.

\section{Physical Origin of differences}

In order to fully appreciate  the nature of the quantitative improvement given by 
our new model,  we will now compare it with the previous ones.

An obvious difference between the model given by Eq.(\ref {eq:H3}) and those 
described by equations
(1) and (\ref {eq:H2}) is that  the former takes into account spin effects. The Monte Carlo
wavefunctions with which we have tested the model were in fact  spin-averaged, but 
$H_3$ contains in principle a full description of
spin-orbit  effects. What follows is a comparison of the models  at the
spin-averaged level. The MonteCarlo wavefunctions obtained in heavy-light 
simulations are typically obtained by averaging the two upper components of
the light quark propagator on the final time slice. That is why we have performed
the same averaging when computing a meson wavefunction from the new RQM.

Expressing the kinetic part of the Hamiltonian $H_3$ in momentum space, we get,
\begin{equation}
\label{eq:H3kin}
H_{3 \; {\rm kin}} = \frac{1}{L^3} \sum_{\bf q} \tilde{\chi}^{+}({\bf q})  \{
 M({\bf q}) \; \beta + \sum_{k=1}^{3} \alpha_{\bf k}  {\bf Q}_{\bf k}({\bf q}) \}  
\tilde{\chi}({\bf q})
\end{equation}
with
\begin{eqnarray}
\label{eq:H3kinmass}
M({\bf q}) & \equiv & m + r \sum_{k=1}^{3} (1 - \cos q_k )  \\
\label{eq:H3kinQ}
{\bf Q}_{\bf k}({\bf q}) & \equiv & \sin q_k  
\end{eqnarray}

Observing Eq.(\ref {eq:H3kin}) and Eq.(\ref {eq:H2}), we realize that  a meaningful
comparison requires
expressing the Dirac-Wilson Hamiltonian of  Eq.(\ref{eq:H3}) in a representation in which
the kinetic energy acquires the form of the kinetic energy piece of Eq.(\ref {eq:H2}).
In the continuum this representation exists  and is given by the well known free 
Foldy-Wouthuysen (FW) transformation \cite{fw}.  By this we mean a transformation 
where the Dirac field is rotated by  the unitary transformation which  decouples
upper and lower components in the {\em absence} of interactions.  Of course,
the full Foldy-Wouthuysen transformation performs this decoupling  including
the interaction with the external field order by
order in the inverse quark mass. However, we wish to avoid a large mass expansion
for light quarks, and an ``all-orders" version of the FW transformation is not known
explicitly. Nevertheless, the relation between models $H_2$ and $H_3$ can still
be clarified by a partial FW transformation in which upper and lower components
are decoupled in the kinetic term only.  On the lattice  the corresponding 
representation goes along the same lines as in the continuum. We then write

\begin{equation}
\label{eq:H3prime}
H_3 ' = \tilde{\chi}^{+} e^{- i S}e^{i S}  {\cal H}_3 e^{- i S}e^{i S}\tilde{\chi}
\end{equation}
where $e^{i S}$ is a unitary (but nonlocal) operator . In momentum space, if we choose $e^{i S}$
according to (See [4])

\begin{equation}
\label{eq:FWoperator}
\langle {\bf p} | e^{i S} | {\bf q} \rangle = L^3  \delta_{{\bf p},{\bf q}} [
\cos \Theta_{\bf q}  + \frac{\gamma^{i} {\bf Q}_i ({\bf q})}{| {\bf Q}({\bf q})
 |}  \sin \Theta_{\bf q} ]
\end{equation}
where ${\bf Q}_i ({\bf q})$ is given by Eq.(\ref{eq:H3kinQ}), $\gamma^{i}$ are the Dirac
gamma matrices, and
\begin{eqnarray}
\label{eq:costeta}
\cos \Theta_{ \bf q} & \equiv & \frac{1}{ \sqrt{2} } 
\sqrt { 1 + \frac{1}{ \sqrt{ 1 +  \frac {| {\bf Q}({\bf q}) |^2 }{M^2({\bf
 q})} } } }   \\
\label{eq:sineteta}
\sin \Theta_{ \bf q} & \equiv & \frac{1}{ \sqrt{2} } 
\sqrt { 1 - \frac{1}{ \sqrt{ 1 +  \frac {| {\bf Q}({\bf q}) |^2 }{M^2({\bf
 q})} } } }
\end{eqnarray}
After this transformation, the kinetic part of $H_3$ becomes 
\begin{equation}
\label{eq:H3primekin}
H_{3\; \rm kin}^{'} = \frac{1}{L^3} \sum_{\bf p} \Psi^+ ( {\bf p})
\beta  E_{\bf p} \Psi ({\bf p})
\end{equation}
where $E_{\bf p} = \sqrt{ M^2({\bf p}) + \sum_{k = 1}^{3} {\bf Q}_k^2({\bf q})
 }$, with $M({\bf p})$ and ${\bf Q}_k({\bf q})$ given by (\ref{eq:H3kinmass}) and
(\ref{eq:H3kinQ}), and $ \Psi \equiv e^{i S} \chi $. In Ref \cite{kron} , a lattice 
Foldy-Wouthuysen transformation is also being considered.

So now, both models have the same kinetic part and  the difference between them becomes
completely transparent. Namely, while the model of Eq.(\ref {eq:H2}) has (in coordinate space)
a potential energy of the form:

\begin{equation}
\label{eq:H2pot}
H_{2\; {\rm pot}} = \sum_{\vec{x}}{\Psi^{+} (\vec{x}) V(\vec{x}) \Psi (\vec{x})}
\end{equation}
 the  potential energy of the  model $H_3$ becomes after the Foldy-Wouthuysen 
transformation of  Eqs (\ref{eq:H3prime}-\ref {eq:sineteta}):

\begin{eqnarray}
H_{3\; {\rm pot}}^{'}  & =  & {1 \over L^6}  \sum_{\vec{z}}  \Psi^+ (\vec{z}) \{   \sum_{\vec{x},\vec{y}}
\sum_{\vec{p} ,\vec{q}}  e^{i \vec{q} (\vec{z} - \vec{x})} e^{i \vec{p} (\vec{x} - \vec{y})} 
V(\vec{x})   \nonumber \\
  &   & \mbox{}   [  \cos \Theta_{ \bf q} + { \vec{\gamma} \cdot \vec{Q} \over |\vec{Q}| }
 \sin \Theta_{\bf q} ]
[  \cos \Theta_{ \bf p} - { \vec{\gamma} \cdot \vec{P} \over |\vec{P}| }
 \sin \Theta_{\bf p} ] \}    \Psi(\vec{y})  
\label{eq:H3primepot}
\end{eqnarray}
as can be seen simply by expressing the fields $\chi$ and $\chi^+$ in terms of  $\Psi $ and $\Psi^+$ through $\chi = e^{-i S} \Psi $ and $\chi^+ =  \Psi^+ e^{i S}$.

Comparing Eqs (\ref {eq:H2pot}) and (\ref {eq:H3primepot}) and taking into account the definitions of  $ \cos \Theta_{ \bf p}$ and $ \sin \Theta_{\bf p} $ given by
Eqs (\ref {eq:costeta}) and (\ref {eq:sineteta}), we see that
(\ref {eq:H3primepot}) reduces to (\ref {eq:H2pot}) in the $m\rightarrow\infty$ limit, 
in which  $ \cos \Theta_{ \bf p} \rightarrow  1 $ and $ \sin \Theta_{\bf p} \rightarrow  0$
and therefore
\begin{eqnarray}
H_{3\;{\rm pot}}'  &\stackrel{m\rightarrow\infty} \longrightarrow&  {1 \over  L^6}  \sum_{\vec{z}} 
 \Psi^+ (\vec{z})  \left\{  \sum_{\vec{x},\vec{y}}
\sum_{\vec{p} ,\vec{q}}  e^{i \vec{q} (\vec{z} - \vec{x})} e^{i \vec{p} (\vec{x} - \vec{y})} 
V(\vec{x})  \right\}  \Psi(\vec{y})  \nonumber  \\
 & = &  \sum_{\vec{x}}{\Psi^{+} (\vec{x}) V(\vec{x}) \Psi (\vec{x})}
\end{eqnarray}

It is worth looking at the above limit  in more detail. Expanding the product  between brackets
in Eq(\ref {eq:H3primepot}) and remembering that $\gamma^i \gamma^j  = g^{i j} - i \sigma^{i j}$, we obtain
\begin{eqnarray}
\label{eq:H3primepotF123}
H_{3\; {\rm pot}}^{'}  & =  & {1 \over L^6}  \sum_{\vec{z}}  \Psi^+ (\vec{z}) \{   \sum_{\vec{x},\vec{y}}
\sum_{\vec{p} ,\vec{q}}  e^{i \vec{q} (\vec{z} - \vec{x})} e^{i \vec{p} (\vec{x} - \vec{y})} 
V(\vec{x})   \nonumber \\
  &   & \mbox{}   [ F_1({\vec p},{\vec q}) + F_2({\vec p},{\vec q}) +
F_3({\vec p},{\vec q}) ] \}    \Psi(\vec{y})  
\end{eqnarray}
where
\begin{eqnarray}
\label{eq:F1}
F_1({\vec p},{\vec q})  & = &  \cos \Theta_{ \bf q}\cos \Theta_{ \bf p} +
 {{\bf Q} ({\bf q}) \cdot {\bf P} ({\bf p}) \over |{\bf Q} ({\bf q})||{\bf P} ({\bf p})| } 
\sin \Theta_{\bf q}\sin \Theta_{\bf p}  \\
\label{eq:F2}
F_2({\vec p},{\vec q}) & = & i  \sigma^{i j} {{\bf Q}_i ({\bf q}){\bf P}_j ({\bf p})
\over |{\bf Q} ({\bf q})||{\bf P} ({\bf p})| } \sin \Theta_{\bf q}\sin \Theta_{\bf p}   \\
\label{eq:F3}
F_3({\vec p},{\vec q}) & = & \gamma^i {{\bf Q}_i ({\bf q}) \over |{\bf Q} ({\bf q})|}
\sin \Theta_{\bf q}\cos \Theta_{ \bf p} -
\gamma^i {{\bf P}_i ({\bf p}) \over |{\bf P} ({\bf p})|}
\sin \Theta_{\bf p}\cos \Theta_{ \bf q} 
\end{eqnarray}

To interpret these terms it is convenient to consider their continuum limit.
In this limit, Eq(\ref {eq:costeta}) and (\ref {eq:sineteta}) become: 
\begin{eqnarray}
\cos \Theta_{ \bf q} & \equiv & \frac{1}{ \sqrt{2} } 
\sqrt { 1 + \frac{1}{ \sqrt{ 1 +  \frac {| {\bf q} |^2 }{m^2} } } }   
\stackrel{m\rightarrow {\rm large}} \longrightarrow     1 - { 1 \over 8}  \frac {| {\bf q} |^2 }{m^2}   \\
\sin \Theta_{ \bf q} & \equiv & \frac{1}{ \sqrt{2} } 
\sqrt { 1 - \frac{1}{ \sqrt{ 1 +  \frac {| {\bf q} |^2 }{m^2} } } }
\stackrel{m\rightarrow {\rm large}} \longrightarrow    { 1 \over 2}  \frac {| {\bf q} | }{m} 
\end{eqnarray}
and  the interactions corresponding to the three terms above become:

\begin{eqnarray}
H_{3\; {\rm pot}}^{' \; F_1}  & =  &   \int_{\vec{z}}  \Psi^+ (\vec{z}) \{   \int_{\vec{x},\vec{y}}
\int_{\vec{p} ,\vec{q}}  e^{i \vec{q} (\vec{z} - \vec{x})} e^{i \vec{p} (\vec{x} - \vec{y})} 
V(\vec{x})   \nonumber \\
\label{eq:H3primeF1}
&   & \mbox{}   [ 1 - { 1 \over 8}  \frac {| {\bf q} |^2 }{m^2} 
- { 1 \over 8}  \frac {| {\bf p} |^2 }{m^2}   + { 1 \over 4}  \frac { {\bf q} \cdot {\bf p} }{m^2}    ] \}
   \Psi(\vec{y})     \\
\label{eq:Coulomb}
& = &  \int_{\vec{x}}  \Psi^+ (\vec{x}) V(\vec{x}) \Psi(\vec{x})     \\
\label{eq:Darwin}
& + &  \frac{1}{8 m^2} \int_{\vec{x}}  \Psi^+ (\vec{x}) \nabla^2 V(\vec{x}) \Psi(\vec{x})   
\end{eqnarray}
Term (\ref {eq:Coulomb}) represents the electrostatic energy of a point-like particle 
and is the one present in models $H_1$ and $H_2$. More interestingly, term 
(\ref {eq:Darwin}) corresponds to exactly the 
Darwin term.  It arises  because of  Zitterbewegung, as can be seen from  the  smearing
of the potential (see Eqs(\ref {eq:H3primepotF123}) and (\ref {eq:F1}) ).

\begin{eqnarray}
H_{3\; {\rm pot}}^{' \; F_2}  & =  &   \int_{\vec{z}}  \Psi^+ (\vec{z}) \{   \int_{\vec{x},\vec{y}}
\int_{\vec{p} ,\vec{q}}  e^{i \vec{q} (\vec{z} - \vec{x})} e^{i \vec{p} (\vec{x} - \vec{y})} 
V(\vec{x})   \nonumber \\
\label{eq:H3primeF2}
&   & \mbox{}   [  i  \sigma^{i j}   \frac{ {\bf q}_i  {\bf p}_j } {| {\bf q}| | {\bf p} |}{ 1 \over 4} 
 \frac {| {\bf q} | | {\bf p} |}{m^2}    ] \}   \Psi(\vec{y})     \\
& = & \int_{\vec{z}}   \Psi^+ (\vec{z}) \{   \int_{\vec{x},\vec{y}}
\int_{\vec{p} ,\vec{q}}  e^{i \vec{q} (\vec{z} - \vec{x})} e^{i \vec{p} (\vec{x} - \vec{y})} 
V(\vec{x})   \nonumber \\
&   & \mbox{}   [  i  \epsilon_{ijk}  \pmatrix{
\sigma^k & 0 \cr
0 & \sigma^k \cr
} \frac{ {\bf q}_i  {\bf p}_j } {4 m^2}    ] \}   \Psi(\vec{y})     \\
\label{eq:spinorbit}
& = &  - \frac{i}{4 m^2} \int_{\vec x}  \Psi^+ (\vec{x})  \epsilon_{ijk}
 {\partial \over \partial x_i}\{{{V({\vec x})}}
\pmatrix{
\sigma^k & 0 \cr
0 & \sigma^k \cr
}  {\partial \over \partial x_j} \Psi(\vec{x}) \}
\end{eqnarray}
Where we have used the identity $\sigma^{i j} = \epsilon_{ijk}  \pmatrix{
\sigma^k & 0 \cr  0 & \sigma^k \cr } $, where the $\sigma^k$  are  the Pauli
matrices. Clearly Eq(\ref {eq:spinorbit}) represents  the spin-orbit interaction.

Finally we have:

\begin{eqnarray}
H_{3\; {\rm pot}}^{' \; F_3}  & =  &   \int_{\vec{z}}  \Psi^+ (\vec{z}) \{   \int_{\vec{x},\vec{y}}
\int_{\vec{p} ,\vec{q}}  e^{i \vec{q} (\vec{z} - \vec{x})} e^{i \vec{p} (\vec{x} - \vec{y})} 
V(\vec{x})   \nonumber \\
\label{eq:H3primeF3}
&   & \mbox{}   [  \gamma^i   \frac{ {\bf q}_i } { 2 m} -    \gamma^i   \frac{ {\bf p}_i } { 2 m } ]  \}   \Psi(\vec{y})   \\
\label{eq:partantipart}
& = &  \frac{i}{2m}  \int_{\vec{x}}  \Psi^+ (\vec{x}) \{ {\partial \over \partial x^i}{V(\vec{x})}
 \gamma^i \}  \Psi(\vec{x})
\end{eqnarray}
representing interactions between upper and lower components of the Dirac spinor.
In a large mass expansion (the usual FW transformation) this term is removed by
a unitary rotation at order $1/m^2$. 

It is perhaps worth mentioning that  ignoring this term 
completely (clearly valid for large masses only!) but {\em not making the large mass
expansion} in (\ref {eq:H3primeF1}, \ref {eq:H3primeF2}) and spin-averaging the resulting
Hamiltonian, yields a modified potential model which we
have studied and which yields wavefunctions very close to the full Dirac formalism.
This approximation is not very well motivated however, as it seems to involve
a rather inconsistent treatment in terms of a $1/m$ development.

In retrospect we realize that potential model descriptions based on $H_1$ and $H_2$ are somewhat  inconsistent  since,  as we have
just seen, they effectively take  the limit  $\; m\rightarrow \infty$ in the potential part 
while keeping a finite mass in the kinetic part (as first shown in Ref \cite{tony2},
 the full relativistic kinetic energy is essential 
in reproducing the data). The reasonably good agreement between  
$H_1$, $H_2$ and the Monte Carlo data, together with  the inconsistency 
pointed out  above, deserves some comments. The validity of  $H_3$, as clearly stated
in section 3, is based on the assumption that, even though the light quark moves 
fast enough  for relativistic effects to be important, the time scales over which, the string 
connecting the quark to the static source responds to changes in the light quark position, 
are small compared with the time scales relevant for the light quark motion. This makes the
interaction between the light and  the static quark  well described by the energy
which would obtain if the light quark were held fixed.  Models
$H_1$ and $H_2$, effectively taking  the limit  $\; m\rightarrow \infty$ in the potential part 
and keeping a finite mass in the kinetic part, are simply making the further assumption
that  the confining potential  is  essentially constant  over regions of  size of the order of the
light quark  Compton wave-length. To  see  this implication  we just have to remember
that a Dirac particle does not move along a straight line with constant velocity but
instead carries out  an oscillatory motion (Zitterbewegung) with the speed of light 
(see \cite{fw,Dirac})
centered on a point which does move uniformly. This oscillatory motion is of the
order of the Compton wavelength of the particle. As our light quark moves though
the confining potential, its color charge explores then  the field over a region of  the 
order of its Compton wavelength and this explains the appearance of the Darwin
term and all higher order terms familiar from the F-W transformation.  
However if
over  regions of the order of the Compton wavelength the field is slowly varying,  it
may be reasonable to ignore the smearing effects (formally higher order in $1/m$)
while maintaining the relativistic kinematics in the kinetic term.
 This seems to be the case in our situation in which,
as can be seen from  the reasonable success of models $H_1$ and $H_2$,
taking the $\; m\rightarrow \infty$ limit in the potential part  seems not to be
a very bad thing to do (for example, at larger light quark masses than those studied
here, the agreement between the wavefunctions generated from $H_2$ and
$H_3$ is closer). However were we going to do the same in the kinetic part, we 
would get a non-relativistic model that does a very bad job at reproducing the wave
functions  \cite{tony2}.

In any case, the most important lesson that we learn from the discussion above  
is that  the differences that we saw in the previous  section between the wave 
functions of model 2 and those of model  3 are, as we have just seen, the result 
of  well known effects that  arise when one combines quantum mechanics
and relativity, which model 3 captures (to the extent that the Dirac equation
captures them),  but are ignored in models 1 and 2. These effects are to our knowledge
visible for the first time in the context of  quantitatively measured (in quenched
lattice QCD)  strong interaction wavefunctions.

\section{Conclusions}

We have presented the results for the 1S and 1P wave functions and energy differences
between the 1S and 2S states of  a fully relativistic lattice model of  heavy-light
mesons. These results were compared  with Monte Carlo measurements of
the corresponding quantities and with previous models. 
The results of the comparison  validated the valence quark model as a
good representation of heavy-light mesons, at least for the lattice
sizes tested so far. In particular our fully relativistic model proved
quantitatively as well as qualitatively superior to previous models.
The quantitative improvement represented by our model arose 
simply by comparison with the data. The qualitative one came not
only from the relative transparency of the approximations being
done, clearly stated in the derivation of  the model; a  comparison
of the physical content of the different models revealed that  the
previous ones were somewhat inconsistent in their relative treatment   
of the potential and kinetic terms. It is precisely this comparison that
allows a  physical interpretation of  the quantitative improvements of the
fully relativistic models. As it turned out they can be thought of as  due to Darwin and 
higher order effects (in the language of a Foldy-Wouthuysen treatment)
 arising from the quantum-relativistic delocalization
of the light quark due to Zitterbewegung. It is remarkable that  the 
Monte Carlo simulations of Ref \cite{tony1} are now accurate enough
to capture this phenomenon.

We expect to be able to extend the above results to much larger lattices.
We are currently generalizing this work to treat mesons with  two finite mass
quarks. If  the fully relativistic model continues to be as
quantitatively accurate as the results obtained here suggest
it may turn out to be a very useful tool in the study of  the
spectrum and static properties of charmonium
and  charmed and B-mesons.

\section{Acknowledgement}

It is a pleasure to thank here A. Duncan for invaluable discussions and
contributions without which this paper could not have been done.
I would also like to acknowledge the assistance of the Fermilab group 
(A.Duncan, E.Eichten, and H.Thacker) for  making available
the Monte Carlo data used in this paper. This work was supported in
part by NSF Grant Phy-9322114 and by the US Dept. of Energy,
Grant No. DE-FG02-91ER40685.

\end{document}